\documentclass[runningheads]{llncs}
\usepackage{graphicx}
\usepackage{amsmath}
\usepackage{hyperref}

\usepackage{enumitem}
\setlist[itemize]{noitemsep, topsep=0pt}
\usepackage[most]{tcolorbox}
\usepackage{cleveref}

\usepackage{orcidlink}

\begin{document}

\title{Converter: Enhancing Interoperability in Research Data Management}
\titlerunning{Converter}

\author{Sefika Efeoglu\inst{1,3}{\orcidlink{0000-0002-9232-4840
}} \and
Zongxiong Chen\inst{2}{\orcidlink{0000-0003-2452-0572}} \and
Sonja Schimmler\inst{1,2}{\orcidlink{0000-0002-8786-7250}}\and
Bianca Wentzel\inst{2}{\orcidlink{0000-0002-9218-5676}}
}
\authorrunning{S. Efeoglu et al.}

\institute{Technische Universität Berlin, Berlin, Germany \\
\email{\{sefika.efeoglu, sonja.schimmler\}@tu-berlin.de}\\
\and
Fraunhofer FOKUS, Berlin, Germany\\
\email{\{zongxiong.chen, bianca.wentzel\}@fokus.fraunhofer.de}
\and
Freie Universität Berlin, Berlin, Germany}

\maketitle            
\begin{abstract}
Research Data Management (RDM) is essential in handling and organizing data in the research field. The Berlin Open Science Platform (BOP) serves as a case study that exemplifies the significance of standardization within the Berlin University Alliance (BUA), employing different vocabularies when publishing their data, resulting in data heterogeneity. The meta portals of the NFDI4Cat and the NFDI4DataScience project serve as additional case studies in the context of the NFDI initiative.
To establish consistency among the harvested repositories in the respective systems, this study focuses on developing a novel component, namely the \textit{converter}, that breaks barriers between data collection and various schemas. With the minor modification of the existing Piveau framework, the development of the converter, contributes to enhanced data accessibility, streamlined collaboration, and improved interoperability within the research community. 
\keywords{Research Data Management  \and Interoperability \and DCAT.}
\end{abstract}

\section{Introduction}

Research Data Management (RDM) plays a crucial role in the research field by facilitating the handling and organization of data. 
As the volume of data in research areas continues to expand, it becomes increasingly important to address the challenge of managing diverse metadata formats  across different applications.
One potential solution to this challenge is to standardize the general descriptive metadata into a common format, e.g., the Data Catalog Vocabulary Application Profile~\footnote{DCAT:~\url{https://www.w3.org/TR/vocab-dcat-3/}} (DCAT-AP)~\cite{DCAT_AP_2019}. 

\begin{figure}[htp]
    \centering
    \caption{An overview of three repositories from the institutions of the Berlin University Alliance on the Berlin Open Science Platform.}
    \tcbox{
    \includegraphics[width=0.6\textwidth]{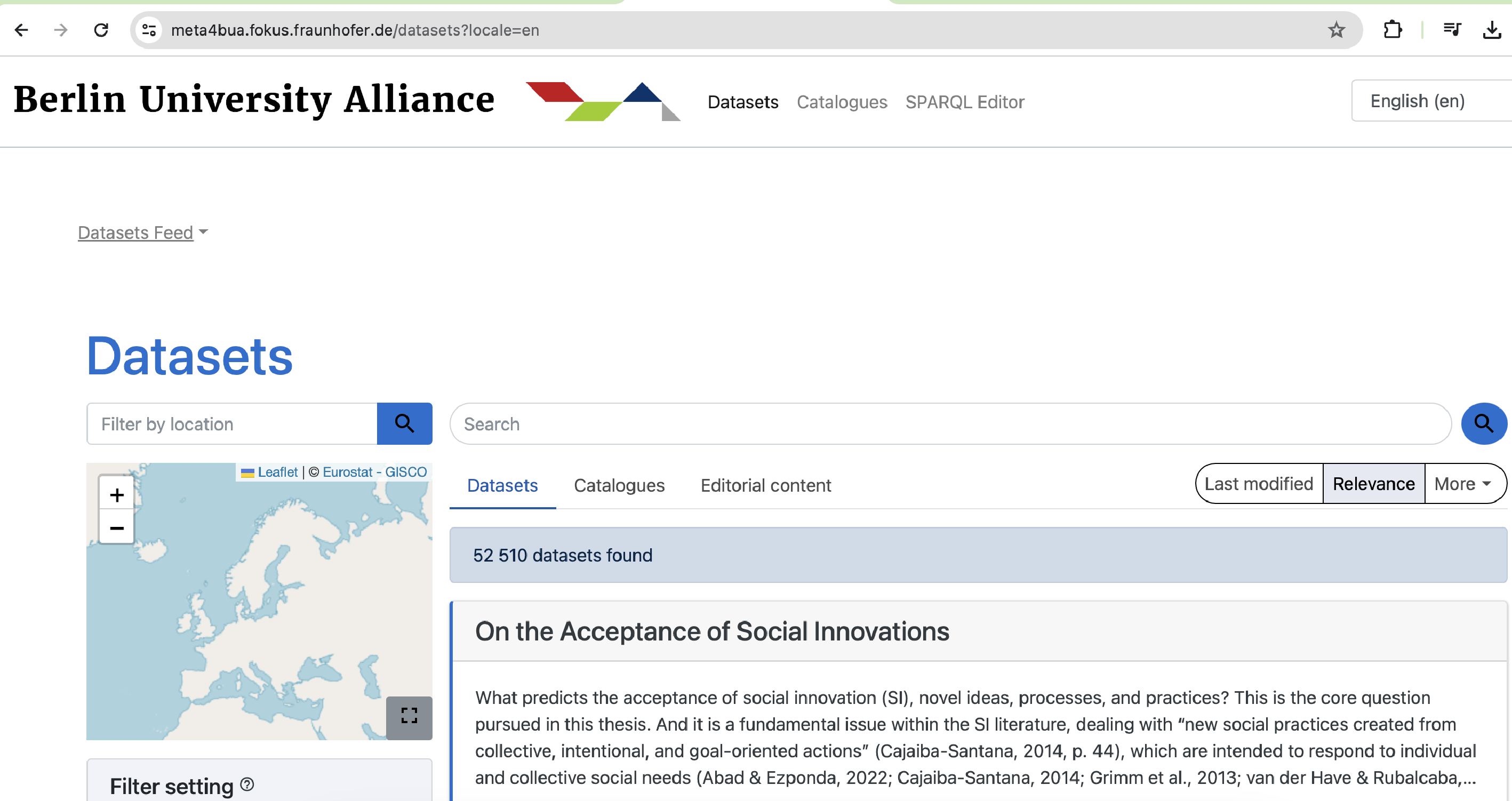}
    }
    \label{fig:bua_platform}
\end{figure}
The Berlin University Alliance (BUA), a German excellency cluster, aims to foster collaboration and knowledge sharing among esteemed institutions: Freie Universität Berlin (FU Berlin), Humboldt-Universität zu Berlin (HU Berlin), Technische Universität Berlin (TU Berlin), and Charité - Universitätsmedizin Berlin~\footnote{This institution's research data is stored in the HU Berlin and FU Berlin repositories}. The BUA's main goal is to establish a digital research space that follows the FAIR (Findable, Accessible, Interoperable, and Reusable) principles, catering to a diverse community of researchers, including professors, scholars, and students. To achieve this, we developed a user-friendly meetup portal called Berlin Open Science Platform (BOP)~\footnote{The metadata portal:~\url{https://meta4bua.fokus.fraunhofer.de/}} (see Figure~\ref{fig:bua_platform}). This platform offers convenient access to data resources from the repositories of three universities, all on a single platform. It is based on Piveau~\footnote{Piveau is available at \url{https://gitlab.com/piveau} and \url{https://github.com/piveau-data}}, a well-established open source data ecosystem, which uses DCAT-AP. The portal brings together and organizes research data from the affiliated universities within the BUA. The research data stored in the BUA repositories encompasses diverse data types including doctoral theses, scientific papers, videos, images, and audio files.

Despite utilizing the oai\_dc format~\footnote{OAI-DC Schema: ~\url{http://www.openarchives.org/OAI/2.0/oai\_dc.xsd}}, the universities within the BUA leverage distinct vocabularies when publishing their data. Using different vocabularies in its schemas causes interoperability problems~\cite{efeoglu2022graphmatcher,LogMap}. The current version of the Piveau harvester requires the adaption of its metadata importer when harvesting data from the different repositories within the BUA. It also lacks the capability to establish the corresponding mapping between DCAT and the data retrieved from the different endpoints. These problems raise the need to convert the schemas of the repositories into DCAT before sending the metadata to the Piveau harvester.

In this work, we developed a novel pipeline that integrates data from different sources and in different schemas. Specifically, we federated data within the BOP in the context of the BUA~\footnote{BUA:~\url{https://www.berlin-university-alliance.de/}}. The approach can be utilized in other projects, including NFDI4Cat~\footnote{NFDI4Cat: ~\url{https://nfdi4cat.org}} and NFDI4DataScience~\footnote{NFDI4DataScience: ~\url{https://www.nfdi4datascience.de}} within the NFDI initiative~\footnote{NFDI:~\url{https://www.nfdi.de}} to build a German National Research Data Infrastructure as well. By employing this pipeline, we can overcome the barriers posed by disparate data resources and harmonize the data into a cohesive and standardized framework. 

In the following, we provide a detailed description of how the proposed converter is implemented and integrated into the Piveau framework. This integration is further extended to its application in the BOP project. The step-by-step process of implementing the converter and integrating it into Piveau is described, while its specific application within the BOP is highlighted. Finally, we summarize the contributions our implementation of the converter makes to the Piveau framework.
\section{Methodology}
\begin{figure}
    \caption{Pipeline: The converter communicates with different repositories and transforms different schemas and vocabularies into a standardized format, i.e., DCAT, and harvester employs importer to fetch metadata from converter and exports to persistent datastore.}
    \label{fig:converter}
    \centering
    \includegraphics[width=0.8\textwidth]{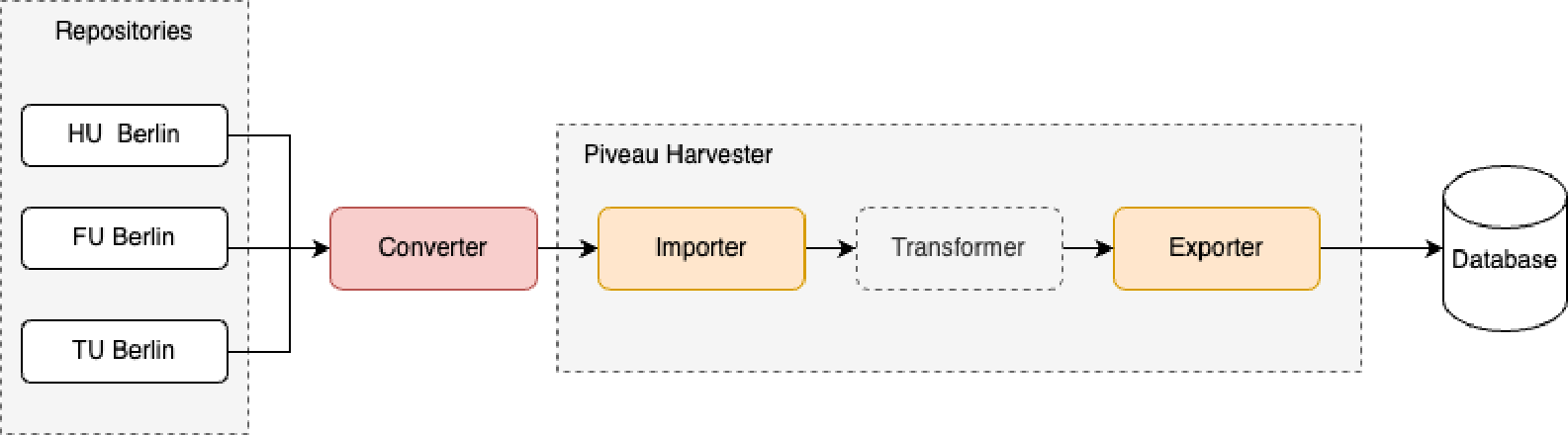}
\end{figure}

We developed a \textit{converter}~\footnote{The converter is available at \url{https://github.com/sefeoglu/dcat-converter}}, which finds the corresponding metadata between the schema of the harvested data and the DCAT vocabulary. The corresponding metadata between schemas is found by a schema matcher (see the GitHub repository). This \textit{converter} facilitates the interoperability between DCAT and data resources using different schemas. After finding the corresponding metadata, it saves the harvesting data in the DCAT format replacing its original schema~\footnote{The sample output of the converter is available at \url{https://github.com/sefeoglu/dcat-converter/blob/master/data/sample.rdf}}. The \textit{converter} acts as a bridge between repositories and the Piveau harvester~\cite{kirstein2019european,piveau_2020} (see Figure~\ref{fig:converter}), offering a set of different importers, transformers and exporters. Despite having its own transformers, the Piveau harvester needs maintenance for various schemas due to receiving metadata in a different format from the DCAT vocabulary. The Piveau harvester in Figure~\ref{fig:converter} exclusively receives data in the DCAT format after integrating the \textit{converter}. As a result, there is no need for additional maintenance in the Piveau harvester, even when the incoming data from endpoints is in different schema formats.
With regards to BOP, an example of the significance of standardization can be observed in the BUA.
This variation poses a challenge in achieving consistency across the harvested repositories within the BOP. The converter transforms the different data formats and schemas into a standardized format, ensuring that data from the DSpace repositories~\footnote{The most commonly used repository solutions include DSpace, Zenodo and Dataverse (see~\url{https://www.re3data.org/}.} of HU Berlin, FU Berlin, and TU Berlin can be easily accessed and utilized, regardless of the specific vocabulary or schema employed by each university. For example, the repositories in the BUA utilize the term ``subject'' in their schemas to define keywords about the data, in contrast to the corresponding term ``keywords'' in DCAT. Another example is that FU Berlin's repository uses ``abstract'' to refer to the publication's abstract, while the other repositories in the BUA use ``description'' in their schemas. We investigated schema alignment among metadata of dcat, oai\_dc, dc\_terms, and dc\_elements, considering their metadata's labels, comments, and definitions, along with prompting ChatGPT by OpenAI and computing cosine similarity with those models' embeddings. We conducted experiments~\footnote{The experiments and their results:~\url{https://github.com/sefeoglu/dcat-converter/tree/master/schema_matching_experiments}} about prompt templates in~\cite{norouzi2023conversational}. Leveraging this tool, BOP can efficiently harmonize and integrate data from all three universities. By addressing the challenge of data heterogeneity, the converter promotes a unified and cohesive research environment.

In the context of the NFDI initiative, we plan to harvest a variety of data repositories in the future. One of the domain-independent repositories we already harvest is the NFDI4Cat Zenodo~\footnote{Zenodo is available at~\url{https://zenodo.org/}} community, which is based on the repository software Invenio. Another repository we harvest in the context of NFDI4Cat is a domain-specific Dataverse instance employed at the BasCat laboratory at TU Berlin. Both do not natively support the \text{DCAT} schema. In order to maintain a cohesive and standardized database, we can integrate new converter services that facilitate the transformation of arbitrary schemas into our targeted \text{DCAT} schema under the proposed framework. By utilizing this service into our data management workflow, we ensure that data from repositories like the Invenio and the Dataverse instance mentioned above can be harmoniously integrated into the NFDI4Cat Meta Portal using the \text{DCAT} format. This conversion process enables consistent data representation and enhances interoperability among the different NFDI projects. The flexibility of the converting script allows for the transformation of varying schemas, accommodating the unique characteristics and structures of different data sources within the research communities. 

\section{Conclusion}
We developed a novel service, \textit{converter}, which resolves the interoperability problem between the DCAT vocabulary and the harvested data before sending the retrieved data to the Piveau framework. Our main contributions are listed in the following. 
\begin{itemize}
    \item Without a converter, the Piveau importer is burdened with the task of managing multiple schemas, necessitating codebase adaption for each distinct schema. Moreover, the transformers currently available in the Piveau framework have limited capabilities in handling complex schema mappings. This limitation poses a challenge in effectively transforming and integrating data from different schemas.
    \item Our work offers a comprehensive solution for effectively managing different vocabularies within the same schema, e.g., oai\_dc. Our proposal stands out due to its seamless deployment and easy integration as a pluggable service into the Piveau framework.
    \item Thanks to the converter introduced, we can eliminate the need for any extensive adaption of the Piveau harvester, streamlining the integration process and ensuring a smooth transition. 
\end{itemize}
\noindent The proposed pluggable service, called \textit{converter}, is initially used to demonstrate how the metadata of the institutes in the Berlin University Alliance (BUA) can be converted into the DCAT format. However, it can be extended to convert the metadata of other universities into the same format within the metadata portal.

\subsubsection*{Acknowledgement.} 
This work has been funded by the German Federal Ministry of Education and Research (BMBF) and the state of Berlin and by the German Research Foundation (DFG) under project numbers 441926934 (NFDI4Cat) and 460234259 (NFDI4DataScience).
 
\bibliographystyle{splncs04}
\bibliography{main}
\newpage
\end{document}